\documentclass[prx,twocolumn,notitlepage,superscriptaddress]{revtex4-2}
\usepackage{amsmath}
\usepackage{amssymb}
\usepackage{graphicx}
\usepackage{dsfont}
\usepackage{comment, color}
\usepackage[caption=false]{subfig}
\usepackage{empheq}

\renewcommand{\k}{\mathbf{k}}
\newcommand{\q}{\mathbf{q}}

\newcommand{\Wph}{ \omega_{\textrm{ph}} }

\newcommand{\Ham}{\hat{H}}

\newcommand{\bra}[1]{\left<#1\right|}
\newcommand{\ket}[1]{\left|#1\right>}
\newcommand{\OPc}[2]{\hat{#1}_{#2}^{\dag}}
\newcommand{\OP}[2]{\hat{#1}_{#2}^{\vphantom{\dag}}}
\newcommand{\CD}[1]{\OPc{c}{#1}}
\newcommand{\C}[1]{\OP{c}{#1}}

\newcommand{\hc}{\textrm{h.c.}}
\newcommand{\E}{\epsilon}

\newcommand{\expect}[1]{\langle #1 \rangle}

\newcommand{\x}{\mathbf{x}}
\renewcommand{\r}{\mathbf{r}}
\newcommand{\rP}{\bar{\mathbf{r}}}
\newcommand{\kP}{\bar{\mathbf{k}}}
\newcommand{\qP}{\bar{\mathbf{q}}}
\newcommand{\p}{\mathbf{p}}

\newcommand{\PhiS}{\hat{\Phi}}
\newcommand{\PsiF}{\hat{\Psi}}

\newcommand{\kk}{\mathds{k}}

\newcommand{\TimeOrder}{\hat{\mathcal{T}}}

\begin{document}
\title{Angle-Resolved Pair Photoemission Theory for Correlated Electrons}
\date{\today}
\author{Thomas P. Devereaux}
\affiliation{Stanford Institute for Materials and Energy Sciences, SLAC National Accelerator Laboratory, 2575 Sand Hill Road, Menlo Park, California 94025, USA}
\affiliation{Geballe Laboratory for Advanced Materials, Stanford University, Stanford, California 94305, USA}
\affiliation{Department of Materials Science and Engineering,
Stanford University, Stanford, California 94305, USA}
\author{Martin Claassen}
\affiliation{Department of Physics and Astronomy, University of Pennsylvania, Philadelphia, PA 19104, USA}
\author{Xu-Xin Huang}
\affiliation{Stanford Institute for Materials and Energy Sciences, SLAC National Accelerator Laboratory, 2575 Sand Hill Road, Menlo Park, California 94025, USA}
\author{Michael Zaletel}
\affiliation{Department of Physics, University of California, Berkeley, California
94720, USA}
\affiliation{Materials Sciences Division, Lawrence Berkeley National Laboratory,
Berkeley, California 94720, USA}
\author{Joel E. Moore}
\affiliation{Department of Physics, University of California, Berkeley, California
94720, USA}
\affiliation{Materials Sciences Division, Lawrence Berkeley National Laboratory,
Berkeley, California 94720, USA}
\author{Dirk Morr}
\affiliation{University of Illinois at Chicago, Chicago, Illinois 60607, USA}
\author{Fahad Mahmood}
\affiliation{University of Illinois at Urbana Champaign, Champaign, Illinois 61801, USA}
\author{Peter Abbamonte}
\affiliation{University of Illinois at Urbana Champaign, Champaign, Illinois 61801, USA}
\author{Zhi-Xun Shen}
\affiliation{Stanford Institute for Materials and Energy Sciences, SLAC National Accelerator Laboratory, 2575 Sand Hill Road, Menlo Park, California 94025, USA}
\affiliation{Geballe Laboratory for Advanced Materials, Stanford University, Stanford, California 94305, USA}
\affiliation{Departments of Physics and Applied Physics, Stanford University, Stanford, CA 94305, USA}

\begin{abstract}
In this paper we consider the possibility and conditions for pair photoemission whereby two incident photons emit pairs of electrons from a candidate material as a novel method to measure and visualize electronic correlations. As opposed to ``double photoemission'' - where a single photon precipitates the ejection of a pair electrons via a subsequent electron energy loss scattering process - we show that pair photoemission need not be limited to interference between initial photoelectrons and valence electrons, and moreover, can occur without the energy penalty of two work functions. This enables detection of pairs of electrons at high energy resolution that may be correlated in the same quantum many-body states. 
\end{abstract}
\maketitle

\section{Introduction}

Over the past decades, angle-resolved photo-emission spectroscopy (ARPES) has emerged as a paradigmatic experimental probe of electronic structure and correlations, band topology or surface states, unconventional superconductivity or the enigmatic pseudogap phase, granting insight to characterize electronic behavior in new quantum materials. By measuring the kinetic energy and angular dependence of photo-emitted electrons, ARPES supplies information on the energy and momentum dependence of valence electrons in a material, and is widely understood to reflect to a good approximation the behavior of the single-particle spectral function \cite{RMP}.

Higher order photoemission processes have been utilized to further obtain information beyond the single-particle density of states. In ``double photoemission'' for example, a highly energetic photon causes the emission of an electron which may cause a second electron to be photoemitted via the Coulomb interaction if it can impart enough energy for the second electron to escape to a detector \cite{Berakdar}. For example, a photoemitted core electron may be accompanied by Auger electron emission, whereby the energy emitted by Auger decay of the core hole is utilized to cause another electron to be emitted \cite{Sawatzky}. Such ``shake-off'' or ``secondaries'' spectra contained both photoemitted core and Auger electrons. The fact that the energies of the two electrons can be themselves continuous yet sum to conserve energy can show that the electrons are correlated, and a comparison with single-particle photoemission can be utilized to determine the so-called ``exchange-correlation'' hole energy \cite{Kirshner}.

In analogy to photon- or electron-based coincidence spectroscopies, recently an interesting proposal suggested extending ARPES to use energy and angle-resolved coincidence detection to account for two-photon two-electron photo-emission events and extract two-particle Bethe-Salpeter wave functions \cite{su20} of valence electrons of the material. Here, in contrast to double photoemission due to Coulomb drag, the coincidence signal derives from two-photon absorption at lower photon energies. Measurement of the angle dependence of two-photon coincidence events at the detector hence importantly permits resolving the momenta and energies of the ejected electron pairs, without the need to ``disentangle'' highly-complex ``Coulomb drag'' processes that complicates the study of important low energy effects.

While the possibility to extract electronic correlations from coincidence counts in ARPES immediately suggests a variety of  applications such as elucidating unconventional pairing mechanisms in high-temperature superconductors or heavy-fermion compounds, a key question concerns understanding exactly the nature of what this new probe actually measures in a correlated electron system, and how to interpret its result in terms of more intuitive quantities such as pair correlation functions or the superconducting gap. Indeed, it is straightforward to see that the coincidence signal does not map onto more readily interpretable superconducting pair correlation functions, since pairs of detected electrons that comprise a coincidence signal are not necessarily ejected from the sample at the same time. On the other hand, a more microscopic description of the pair ARPES cross section is necessary, to permit a formal accounting for final-state effects and the detector geometry and differentiate from double photoemission.

In this work, we address these questions by developing a generic theoretical description of angle-resolved pair photoemission and studying its behavior in light of superconducting instabilities in the attractive and repulsive Hubbard model on small clusters. We show that pair photoemission need not be limited to interference between initial photoelectrons and valence electrons, and moreover, can occur without the energy penalty of two work functions. This enables detection of pairs of electrons at high energy resolution that may be correlated in the same quantum many-body states. 

\begin{figure}[t]
	\centering
	\includegraphics[width=\columnwidth,clip]{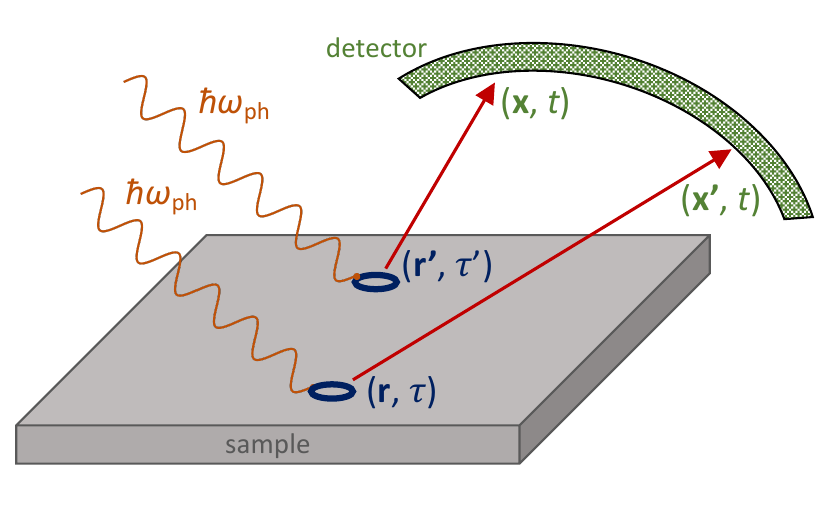}
	\caption{\textbf{Setup: } Two electrons at $(\rP,\tau)$, $(\rP',\tau')$ are photo-emitted upon absorbing two photons with frequency $\Wph$. Coincidence detection takes place at time $t$ and positions $\x$, $\x'$.} \label{fig:setup}
\end{figure}

Fig. \ref{fig:setup} depicts a schematic of the pair photoemission process for a two-dimensional sample or surface state. Two photons with energy $\hbar\Wph$ eject two electrons from the sample, which are subsequently observed at the detector at the same time $t$ with both angle and energy resolution. For simplicity, but without loss of generality we ignore bulk effects, and henceforth denote three-dimensional positions and momenta using bold notation $\r$, whereas their two-dimensional components in the sample plane are denoted by $\rP$. Suppose that sample and emitted electrons are described by fields $\PhiS(\rP)$ and $\PsiF(\r)$, respectively (we suppress implicit spin indices, for conciseness), and are governed by a generic Hamiltonian $\Ham$
\begin{align}
	\Ham_0 &= \Ham_{\textrm{valence}}(\PhiS) + \Ham_{\textrm{emitted}}(\PsiF) + \Ham_{\textrm{v-e}}(\PhiS, \PsiF)   \label{eq:H0}
\end{align}
such that emitted electrons $\PsiF$ behave as freely-propagating waves at long distances from the sample while appropriately encapsulating final-state effects (inverse LEED) as well as possible back actions $\Ham_{\textrm{v-e}}(\PhiS, \PsiF)$ which would be important for Coulomb-drag mediated double photoemission.

The photoemission process $\Ham_{\textrm{el-ph}}$ now takes a sample electron $\PhiS(\rP)$ to a propagating final state $\PsiF(\r)$
\begin{align}
	\Ham_{\textrm{el-ph}}(t) &= s(t) \int d^3\r~ g(\r)~ \PsiF^\dag(\r) \PhiS_S(\rP)~ e^{-i \Wph t} + \hc
\end{align}
where $g(\r)$ is the dipole matrix element and $s(t)$ describes a Gaussian probe pulse envelope with
\begin{align}
	s(t) = e^{-(t-t_0)^2/2\sigma_{\textrm{pr}}^2 }
\end{align}
Subsequently, the photo-electron detector measures the mean momentum $\kk$ of propagating electron wave packets, described by a photo-current
\begin{align}
	\expect{ \hat{J}_{\kk} } = \frac{\kk}{e} \iint d\x d\x'~ \phi_{\kk}^\star(\x) \phi_{\kk}(\x') ~\expect{ \PsiF^\dag(\x) \PsiF(\x') }
\end{align}
where $\phi_{\kk}(\r)$ denotes a wave packet centered at the detector location.

\section{Formalism}

\subsection{Single-Electron ARPES}

The conventional ``single-electron'' ARPES signal now follows straightforwardly \cite{freericks09} from a perturbative expansion in $\Ham_{\textrm{el-ph}}$ of the measured photocurrent
\begin{align}
	I_{\kk} &= \int_{-\infty}^t d\tau d\tau' s(\tau) s(\tau') e^{i\Wph(\tau-\tau')} \notag\\
    &\times \int d\x d\x' \phi_{\kk}^\star(\x) \phi_{\kk}(\x') \int d\rP d\rP' g^\star(\r) g(\r') \notag\\	
    &\times \expect{ \PhiS^\dag(\rP,\tau) \PsiF(\r,\tau) \PsiF^\dag(\x,t) \PsiF(\x',t) \PsiF^\dag(\r',\tau') \PhiS(\rP',\tau') }
\end{align}
where $\langle \cdot \rangle = \textrm{tr} \{ \cdot ~ e^{-\beta \Ham_0} \} / Z$ denotes thermal expectation values with respect to $\Ham_0$. If back action $\Ham_{\textrm{valence-emitted}}$ between emitted and valence electrons can be neglected, this expression simplifies drastically, as $\langle \PhiS^\dag(\rP,\tau) \PsiF(\r,\tau) \PsiF^\dag(\x,t) \PsiF(\x',t) \PsiF^\dag(\r',\tau') \PhiS(\rP',\tau') \rangle = \langle \PhiS^\dag(\rP,\tau) \PhiS(\rP',\tau') \rangle \langle \PsiF(\r,\tau) \PsiF^\dag(\x,t) \rangle \langle \PsiF(\x',t) \PsiF^\dag(\r',\tau') \rangle$. Furthermore, assuming a single electronic valence band $\PhiS(\rP) = \sum_{\kP} u_{\kP}(\rP) e^{i\kP \rP} \C{\kP}$ with Bloch function $u_{\kP}(\rP)$, and neglecting the detector wave packet shape functions $\phi_{\kk}(\x) \to e^{i\kk \x}$ (thereby discarding time-of-flight information), one arrives at ($\kk \to \k$):
\begin{align}
	I_{\k} = -i \int_{-\infty}^t d\tau d\tau' & e^{i\Wph(\tau-\tau')}  s(\tau) s(\tau') \left| M_{\k} \right|^2 \times \notag\\
		& \times~ G^<_{\kP}(\tau,\tau') \mathcal{G}^\star_{\k}(\tau,t) \mathcal{G}_{\k}(t,\tau')
\end{align}
where $M_{\k}$ is a matrix element evaluated from $g(\r)$ and the Bloch function of the single valence band, $G^<_{\kP}(\tau,\tau')$ is the lesser sample Green's function
\begin{align}
	G^<_{\kP}(\tau,\tau') = i \expect{ \CD{\kP}(\tau) \C{\kP}(\tau') }
\end{align}
and
\begin{align}
	\mathcal{G}_{\k}(t,t') = -i \expect{ \TimeOrder\C{\k}(t) \CD{\k}(t') }
\end{align}
is the propagating electron Green's function for an inverse LEED state. Finally, a drastically simplified expression can be provided, if $\mathcal{G}_{\k}$ is approximated by a free electron Green's function with dispersion $\E_\k = \k^2/2m_0$. Then, defining the energy $\omega$ observed at the detector as
\begin{align}
	\omega \equiv \Wph - \frac{\k^2}{2m_0} - W
\end{align}
where $W$ is the work function of the sample, and taking $t \to \infty$, one finally arrives at
\begin{align}
	I_{\k} = i \int_{-\infty}^{\infty} d\tau d\tau' e^{i\omega(\tau-\tau')} s(\tau) s(\tau') \left| M_{\k} \right|^2 G^<_{\kP}(\tau,\tau') 
\end{align}
This is the usual expression for single-particle ARPES in terms of convolutions of shape functions, matrix elements and the lesser Green's function\cite{freericks09}.

\subsection{Angle-resolved Pair Photoemission}

Similarly, a coincidence measurement signal can be defined as
\begin{align}
	&\langle \hat{J}_{\kk_1\kk_2} \rangle = \frac{\kk_1\kk_2}{e^2} \int d\x_1 d\x_1' d\x_2 d\x_2'~ \phi_{\kk_1}^\star(\x_1) \phi_{\kk_2}^\star(\x'_1) \phi_{\kk_2}(\x'_2)  \notag\\
		&~~~~\times \sum_{\nu\nu'} \phi_{\kk_1}(\x_2)~ \expect{ \PsiF_{\nu}^\dag(\x_1,t) \PsiF_{\nu'}^\dag(\x'_1,t) \PsiF_{\nu'}(\x'_2,t) \PsiF_{\nu}(\x_2,t) }
\end{align}
In complete analogy to single-electron ARPES, the photo-detection rate can now be evaluated from a perturbative expansion in $\Ham_{\textrm{el-ph}}$. To first order the response involves a single photoemission vertex and vanishes.
We note that this contribution is essential for Coulomb-mediated double photoemission for high photon energies. Here, a perturbative expansion in $\Ham_{\textrm{v-e}}(\PhiS, \PsiF)$ additionally accounts for the Coulomb interaction mediated back action of the photo emitted electron, imparting enough energy on a second sample electron to eject it, rendering the coincidence signal non-zero. As discussed above, we are primarily interested in two-photon two-electron pair ARPES processes at lower photon energy; in this regime, the double emission contribution is negligible for energetic reasons.

To second order in $\Ham_{\textrm{el-ph}}$, the two-photon two-electron coincidence photo-detection signal formally reads
\begin{align}
    D_{\kk_1\kk_2} &= \sum_{\substack{\sigma_1\sigma'_1\nu \\ \sigma_2\sigma'_2\nu'}}~ \int\displaylimits_{-\infty}^t d\tau_1 d\tau_2 \int\displaylimits_{-\infty}^{\tau_1} d\tau'_1 \int\displaylimits_{-\infty}^{\tau_2} d\tau'_2 \notag\\
    &\times e^{i\Wph(\tau_1 + \tau'_1 - \tau_2 - \tau'_2)} s(\tau_1) s(\tau'_1) s(\tau_2) s(\tau'_2)  \notag\\
    &\times \int d\r_1 d\r'_1 d\r_2 d\r'_2 g^\star(\r_1) g^\star(\r'_1)  g(\r'_2) g(\r_2) \notag\\
    &\times \int d\x_1 d\x'_1 d\x_2 d\x'_2~ \phi_{\kk_1}^\star(\x_1) \phi_{\kk_2}^\star(\x'_1) \phi_{\kk_2}(\x'_2) \phi_{\kk_1}(\x_2)  \notag\\
    &\times \langle \left< \PhiS_{\sigma_1}^\dag(\r_1,\tau_1) \PsiF_{\sigma_1}(\r_1,\tau_1) \PhiS_{\sigma'_1}^\dag(\r'_1,\tau'_1) \PsiF_{\sigma'_1}(\r'_1,\tau'_1) \right. \notag\\
        &~~~\times \PsiF_{\nu}^\dag(\x_1,t) \PsiF_{\nu'}^\dag(\x'_1,t) \PsiF_{\nu'}(\x'_2,t)  \PsiF_{\nu}(\x_2,t) \notag\\
        &~~~\times\left. \PsiF_{\sigma'_2}^\dag(\r'_2, \tau'_2) \PhiS_{\sigma'_2}(\r'_2,\tau'_2) \PsiF_{\sigma_2}^\dag(\r_2, \tau_2) \PhiS_{\sigma_2}(\r_2,\tau_2) \right>
    &
\end{align}
Assuming negligible back action or Coulomb interactions between photo-emitted electrons and low-energy sample electrons, this daunting multi-point correlation function can be decomposed in analogy to conventional ARPES. The coincidence detection rate can be written as
\begin{align}
&D_{\kk_1\kk_2} = \int\displaylimits_{-\infty}^t d\tau_1 d\tau_2 \int\displaylimits_{-\infty}^{\tau_1} d\tau'_1 \int\displaylimits_{-\infty}^{\tau_2} d\tau'_2~  s(\tau_1) s(\tau'_1) s(\tau_2) s(\tau'_2) \notag\\
    &\times \int d\k d\k' d\q \sum_{\substack{\sigma_1\sigma'_1\nu \\ \sigma_2\sigma'_2\nu'}} G_{\kP\kP'\qP}^{\substack{\sigma_1\sigma_2 \\ \sigma'_1\sigma'_2}}(\tau_1,\tau'_1,\tau'_2,\tau_2) e^{i\Wph(\tau_1 + \tau'_1 - \tau_2 - \tau'_2)} \notag\\
    &\times   \left[ F_{\k\q}^{\sigma_1\sigma'_1\nu\nu'}(\tau_1,\tau'_1) \right]^* F_{\k'\q}^{\sigma_2\sigma'_2\nu\nu'}(\tau_2,\tau'_2)
\end{align}
where
\begin{equation}
	G_{\kP\kP'\qP}^{\substack{\sigma\nu \\ \sigma'\nu'}}(\tau_1,\tau'_1,\tau'_2,\tau_2) = \expect{ \CD{\kP\sigma}(\tau_1) \CD{\qP-\kP\sigma'}(\tau'_1) \C{\qP-\kP'\nu'}(\tau'_2) \C{\kP'\nu}(\tau_2) }
\end{equation}
is the two-particle Green's function for the sample, and
\begin{align}
	F_{\k\q}^{\sigma\sigma'\nu\nu'}(\tau,\tau') &= \int d\p d\p' ~\phi_{\kk_1}(\p) \phi_{\kk_2}(\q-\p)~ M_{\q-\k} M_{\k} \notag\\
	 &\times \bra{0} \PsiF_{\p\nu'}(t) \PsiF_{\q-\p\nu}(t) \PsiF^\dag_{\q-\k\sigma'}(\tau') \PsiF^\dag_{\k\sigma}(\tau)  \ket{0}  \label{eq:PhotoEmittedCorrelator}
\end{align}
is a four-point function for the photo-emitted electrons which includes the Fourier-transformed detector shape functions $\phi_{\kk}(\cdot)$ and evaluated with respect to the vacuum state. Furthermore, $M_{\k}$ denote photo-excitation matrix elements defined in terms of the dipole matrix element and valence electron Bloch functions introduced above. $F_{\k\q}(\tau,\tau')$ encodes both propagation and time-of-flight information, as well as interactions between the two photo-emitted electrons. A drastic simplification follows from treating emitted electrons as free fermions. In this case, $F_{\k\q}(\tau,\tau')$ factorizes to
\begin{align}
	&F_{\k\q}(\tau,\tau') = M_{\q-\k} M_{\k}~ e^{-i[ \E_{\k}(t-\tau) + \E_{\q-\k}(t-\tau')]}  \notag\\ 
	&\times \left[ \phi_{\kk_1}(\k) \phi_{\kk_2}(\q-\k) \delta_{\sigma,\nu'} \delta_{\sigma',\nu}  - \phi_{\kk_2}(\k) \phi_{\kk_1}(\q-\k) \delta_{\sigma,\nu} \delta_{\sigma,\nu} \right]
\end{align}
where $\E_\k = \k^2/2m_0$ is the dispersion of the photo-emitted electrons.
In analogy to the theory for conventional ARPES \cite{freericks09}, we can now make the assumption that the detector wave packet momentum width can be neglected $\phi_{\kk}(\k) \to \delta(\kk - \k)$, discarding again time-of-flight information and dependence on the detector position. Denote the energies observed at the two detectors minus the photon energy as
\begin{align}
	\omega_{1,2} \equiv \Wph - \frac{\kk_{1,2}^2}{2 m_0} - W
\end{align}
with $W$ the work function of the sample, and taking $t \to \infty$, the coincidence detection rate can be written as
\begin{align}
	D^{(0)}_{\k_1\k_2} &= \int\displaylimits_{-\infty}^{\infty} d\tau_1 d\tau'_1 d\tau_2 d\tau'_2~  s(\tau_1) s(\tau'_1) s(\tau_2) s(\tau'_2)  \notag\\
 &\times \sum_{\sigma\sigma'} \expect{ \TimeOrder \CD{\kP_1\sigma}(\tau_1) \CD{\kP_2\sigma'}(\tau'_1) \TimeOrder \C{\kP_2\sigma'}(\tau'_2) \C{\kP_1\sigma}(\tau_2) } \notag\\
 &\times e^{i \left[ \omega_1(\tau_1 - \tau_2) + \omega_2 (\tau'_1 - \tau'_2)\right]}\label{eq:CARPES}
\end{align}
where $\TimeOrder$ denotes time ordering, and we additionally omitted the photo-excitation matrix elements $M_\k$ for conciseness. 

\subsection{Fermi's Golden Rule}

This expression can be recast in a more familiar Fermi's Golden rule by inserting complete sets of the states for the $N, N-1,$ and $N-2$ particle sectors. Also if we neglect the time dependence of the shape functions so that we only concentrate on frequency resolution, the time integrals can be performed and the following expression is obtained:
\begin{align}
    D^{(0)}_{\k_1,\k_2}(\omega_1,\omega_2) &= \sum_n \mid M_{0,n}(\k_1,\k_2,\omega_1,\omega_2)\mid^2  \notag\\ 
&\times \delta(E_n-E_0+\omega_1+\omega_2)
    \label{Eq:p-arpes1}
\end{align}
with $E_0$ denoting the $N$ particle ground state energy and $E_n$ the eigenenergies of the $N-2$ particle sector - viz., the expression is simply a matrix element squared times a term that enforces energy conservation. The matrix element reads 
\begin{align}
    M_{0,n}(\k_1,\k_2,\omega_1,\omega_2) = &\sum_{m,\sigma_1,\sigma_2} \left\{ \frac{ \bra{n}{\C{\k_2\sigma_2}}\ket{m} \bra{m}{\C{\k_1\sigma_1}}\ket{0}}
    {E_m-E_0+\omega_1 - i\eta}  
    \right. \notag\\
    -& \left.  \frac{ \bra{n} \C{\k_1\sigma_1} \ket{m} \bra{m} \C{\k_2\sigma_2} \ket{0}}
    {E_m-E_0+\omega_2 - i\eta}
\right\}
    \label{Eq:p-arpes2}
\end{align}
with $E_m$ the $N-1$ particle sector eigenvalues.

Note that this expression bears a strong resemblance to the Kramers-Heisenberg expression for resonant inelastic x-ray scattering (RIXS) in which the manifold of $N-1$ states $\{\mid m\rangle\langle m \mid \}$ play the role of intermediate $N+1$ core hole states whereby a core electron is photoexcited into the valence band\cite{RIXS}. While for RIXS the final states have the same number of electrons $N$ as the initial state as the core hole is refilled via photo-deexcitation, the pair photoemission final states have two less electrons $N-2$. Despite their apparent differences, the functional form of Eq. (\ref{Eq:p-arpes2}) indicates that we would expect resonant pair photoemission whenever one or both of the frequencies $\omega_{1,2}$ correspond to the $N-1$ removal state energies observed in photoemission rather than the core-valence transition energies as in RIXS.

To illustrate the differences between pair photoemission and uncorrelated single particle photoemission and how information can be obtained from both, we start by reminding that the "pairing energy" $\Delta({\bf k_1, k_2})$ for two momentum states ${\k_{1,2}}$ is $\Delta({\bf k_1, k_2})= E_2({\bf k_1,k_2}) - E_1({\bf k_1})-E_1({\bf k_2})+E_0$ where $E_N$ denotes the energies of the $N$ particle removal states, viz., where single particle photoemission yielding $E_1-E_0$ and pair photoemission $E_2-E_0$. By inspection of Eqs. (\ref{Eq:p-arpes1}) and (\ref{Eq:p-arpes2}), we can see that $E_2-E_0$ is determined by the overall energy conservation by $\omega_1+\omega_2$. In other words, this is given by the slope of the line connecting $\omega_1$ and $\omega_2$ for pair photoemission when plotted as a function of both frequencies. The resonance denominator of Eq. (\ref{Eq:p-arpes2}) shows that the intensity on this line is modulated when $\omega_{1,2}$ coincide with the single particle energies $E_1-E_0$ observed in single particle photoemission.


\subsection{Retarded pairing correlator}

Suppose that the measured pair emission signal is obtained as a function of sum and difference frequencies
\begin{align}
	\omega = \omega_1 + \omega_2~,~~~ \Delta\omega = \omega_1 - \omega_2
\end{align}
By inspection of Eq. (\ref{eq:CARPES}) one can see that the difference frequency $\Delta\omega$ parameterizes the ``retardation'' of the pair emission process from the sample, i.e. the time delay between emission of the first and second electron of an observed pair. Integration over the difference frequency $\Delta\omega$ then yields
\begin{align}
	D^{(0)}_{\k_1\k_2}(\omega) &= \int\displaylimits_{-\infty}^{\infty} d\Delta\omega~ D^{(0)}_{\k_1\k_2}\left(\frac{\omega + \Delta\omega}{2}, \frac{\omega - \Delta\omega}{2}\right) \notag\\
	= \int\displaylimits_{-\infty}^{\infty} dt& ~e^{-i\omega t} \int\displaylimits_{-\infty}^{\infty} d\tau \expect{ \CD{\k_1}(t) \CD{\k_2}(t+\tau) \C{\k_1}(\tau) \C{\k_1}(0) }
\end{align}
which can be straightforwardly expressed as a spectral decomposition
\begin{align}
	D^{(0)}_{\k_1\k_2}(\omega) &= \sum_{n,m}& \delta(\omega + E_0 - E_n) \left| \bra{n} \C{\k_1} \ket{m} \bra{m} \C{\k_2} \ket{0} \right|^2
	\label{Eq:pfc}
\end{align}

Thus coincidence pair ARPES can yield the dynamic superconducting pairing susceptibility. In a BCS superconductor, the resulting response has a peak at finite $\omega$ that corresponds to the momentum-dependent superconducting gap. Spin-resolved pairing as well as pairing that can occur at finite momenta corresponding to a pair density wave was recently examined in Ref. \onlinecite{Fahad}. Importantly, coincidence pair ARPES can also provide measurements for the dynamic pair susceptibility in materials at temperatures above the ordered phase or for systems that may be highly frustrated or condense into a different, non-superconducting pair state. While dynamic pairing correlations have been measured via different numerical methods, such as determinant Quantum Monte Carlo for example \cite{Maier,Frontiers}, susceptibility measurements have been lacking.

\section{Applications}

\subsection{Free electrons}

If the valence electrons within the sample are free, the four-point function factorizes to
\begin{align}
	G_{\k_1,\k_2}^{(2)}(\tau_1,\tau'_1,\tau'_2,\tau_2) \to G^<_{\k_1}(\tau_1 - \tau_2) ~ G^<_{\k_2}(\tau'_1 - \tau'_2)
\end{align}
for $\k_1 \neq \k_2$ and the coincidence detection rate becomes a product of single-particle ARPES detection rates
\begin{align}
	D_{\k_1\k_2}^{(0)}(\omega_1,\omega_2) \to P_{\k_1}(\omega_1) P_{\k_2}(\omega_2)
\end{align}
which only contributes if the quantum numbers of the photodetected electrons are not identical due to Pauli exclusion. This is a useful check to determine the overall magnitude of the pair ARPES compared to single particle ARPES, and can help to assess the spectral intensities of two-particle collective modes separately from the single particle continuum.

\subsection{BCS theory}

If the system of interest is well-described by a BCS mean field ansatz, the valence band is again composed of free Bogoliubov fermions. In this case, the four-point function factorizes, and the coincidence pair photoemission signal additionally includes a pairing term:
\begin{align}
	D_{\k_1\k_2}^{(0)}(\omega_1,\omega_2) = P_{\k_1}(\omega_1) P_{\k_2}(\omega_2) + \left| P_{\k_1\k_2}^{\textrm{pair}}(\omega_1,\omega_2) \right|^2
    \label{Eq:BCS_Full}
\end{align}
where
\begin{align}
	P^{\textrm{pair}}_{\k_1\k_2}(\omega_{1,2}) = \int d\tau d\tau' s(\tau) s(\tau') e^{i(\omega_1 \tau + \omega_2 \tau')} \expect{ \mathcal{T} \CD{\k_1}(\tau) \CD{\k_2}(\tau')}
\end{align}
is a weighted time average of the time-ordered anomalous Green's function. For the case where the shape functions $s(t)=1$ the BCS singlet pair wavefunction gives the value
\begin{align}
	P^{\textrm{pair}}_{\k_1\k_2}(\omega_{1,2}) &= \delta(\k_1+\k_2)\delta(\sigma_1+\sigma_2)
	\delta(\omega_1+\omega_2)\notag\\
	&\times \frac{\Delta_{\k_1}}{(\omega_1-i\eta)^2-E_{\k_1}^2}
	\label{Eq:BCS}
\end{align}
with the Bogoliubov energy given by $E^2_{\k}=\epsilon_{\k}^2+\Delta^2_{\k}$ for free particle dispersion $\epsilon_{\k}$, and $\eta$ is a small real quantity \cite{su20}.

We note that Eq. (\ref{Eq:BCS}) yields a sharp peak at the Fermi level ($\omega_{1,2}=0$) when the delta functions are satisfied, indicating that pair ARPES can be used to detect the underlying Cooper pair structure in terms of center of mass spin (i.e., singlet versus triplet) and momentum (i.e., Fulde-Ferrel or pair density-wave state) as has been noted previously \cite{su20,Fahad}. Moreover, the fermion momentum dependence of the energy gap $\Delta(\k)$ can be scanned and directly measured.

\subsection{Hubbard models for correlated electrons}
\begin{figure*}
    \centering
    \includegraphics[width=\textwidth]{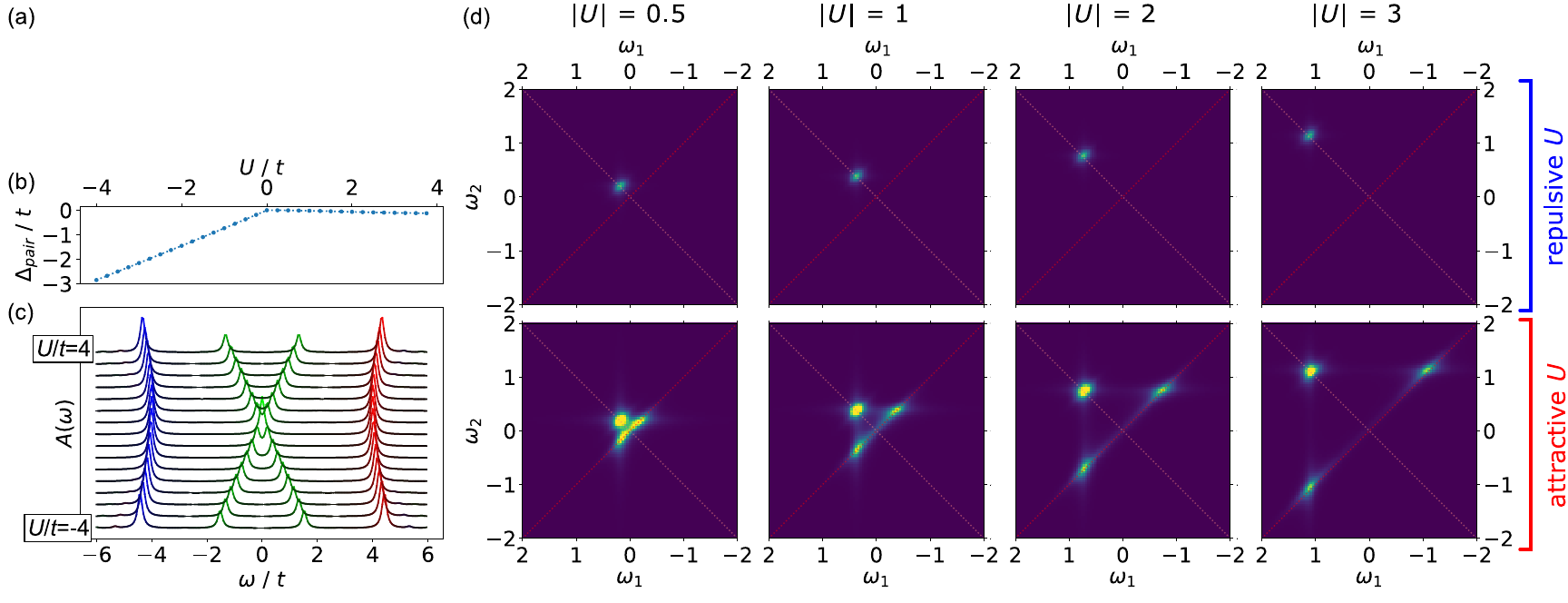}
    \caption{\textbf{Pair ARPES for the attractive and repulsive Hubbard model.} (a) Schematics of the eight-site Betts cluster. (b) Pair binding energy as a function of repulsive and attractive Hubbard interactions at half filling. (c) Single-particle spectrum $A(\omega)$ as a function of interactions and momentum (blue, green, red correspond to $\mathbf{k}=0, (\pi/2,\pi/2), (\pi,\pi)$, respectively). (d)
    Top and bottom row panels show opposite-spin $D_{\mathbf{k},-\mathbf{k}}(\omega_1,\omega_2)$ for repulsive and attractive interactions, respectively, from $|U|=0.5$ (left) to $|U|=3$ (right). Dashed lines (center and difference frequencies) are guides to the eye. All energies are quoted in units of $t=1$. While the main photoemission peak with $\omega_1 = \omega_2$ identically tracks attractive and repulsive Coulomb interactions, for $U < 0$ pair ARPES reveals the pair-breaking intermediate state via the frequency difference spectrum $\omega_1 = \omega_2$ for $\omega_1 + \omega_2 = 0$ (bottom row).}
    \label{fig:PairARPES}
\end{figure*}

The single band Hubbard model may provide a simple way to characterize the behavior of pair photoemission for correlated electrons in systems without superconducting long-range order. Specifically we will utilize eigenstates of the particle-hole symmetric Hubbard model 
\begin{align}
    H = -t\sum_{\langle i,j\rangle,\sigma} c^\dagger_{i,\sigma} c_{j,\sigma} + U \sum_i (n_{i,\uparrow}-\frac{1}{2}) (n_{i,\downarrow}-\frac{1}{2}) 
\end{align}
on an 8A (diamond) Betts cluster \cite{Betts} to construct pair ARPES [Fig. \ref{fig:PairARPES}(a)]. Here $c_{i,\sigma},c^\dagger_{i,\sigma}$ removes, adds a particle at site $i$ with spin $\sigma$, $n_{i,\sigma}$ is the particle density per spin at site $i$, $t$ denotes hybridization between nearest neighbor sites $i$ and $j$, and $U$ is a measure of the local interaction between opposite spins. Throughout we assume units where $\hbar=1$.

While much work has been performed via density matrix group renormalization techniques (DMRG) for example to ascertain whether the Hubbard model in the thermodynamic limit harbors superconductivity, our goal is more modest. By examining the eigenstates and constructing pair ARPES on finite clusters, which cannot have a bona fide phase transition, we may be able to highlight how coincidence spectroscopy can be used to quantitatively measure pair field susceptibilities in systems where $U(1)$ gauge symmetry is not broken but fluctuating order may be inferred.

Pairing has been long investigated in exact diagonalization studies of the Hubbard model on small clusters \cite{Scalapino,Young,Marsiglio}. The pair binding energy $\Delta$ is defined as the energy difference between the ground state energies of $N$ and $N-2$ particle systems minus twice the energy of the $N-1$ system: 
\begin{equation}
    \Delta = E_N+E_{N-2}-2E_{N-1}
\end{equation}
A negative $\Delta$ indicates an effective electron pair attraction. 

The pair binding energy $\Delta$ obtained for the repulsive and attractive Hubbard model at half-filling $N_{electrons}=8=N$ is shown in Fig. \ref{fig:PairARPES}(b). For repulsive $U$, $\Delta$ is negative for $U/t \lesssim 8$ and becomes positive for larger values. 
The ground state of the attractive Hubbard model ($U<0$) can be well modeled as a BCS superconducting paired state \cite{Marsiglio}, and possesses a pair binding energy that increases with $\mid U\mid$. 
                                 
The ARPES spectra are identical for repulsive or attractive $U$ via particle-hole symmetry. Fig. \ref{fig:PairARPES}(b) depicts the spectral functions as a function of $U/t$, with peaks corresponding to the three unique momenta on the 8A Betts cluster $(0,0), (\pi/2,\pi/2)$ (6-fold degenerate), and $(\pi,\pi)$, shown in blue, green and red. As noted previously \cite{Elbio}, spectral peaks move to deeper binding energies as $\mid U\mid$ is increased, and the development of the lower Hubbard band can be more clearly observed. While a pairing gap is clearly observable for attractive $U$, a superconductor cannot be distinguished from a Mott gap in single-electron ARPES. Indeed, the spectra for positive and negative $U$ are identical by virtue of particle hole symmetry. This further motivates investigating pair photoemission, which intrinsically discriminates between pair and density excitations.

\begin{figure*}[t]
    \centering
    \includegraphics[width=\textwidth]{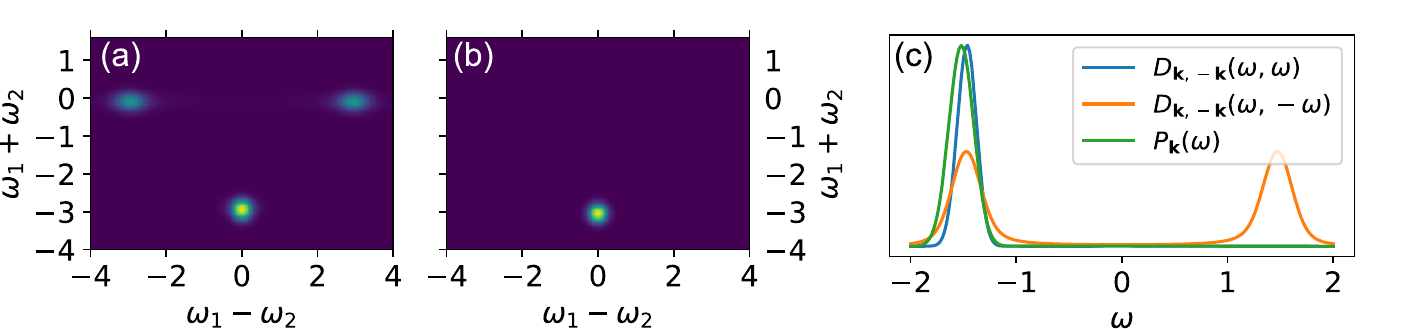}
    \caption{\textbf{Comparing Single and Pair Photoemission.} (a) and (b) depict Pair ARPES $D_{\mathbf{k},-\mathbf{k}}(\omega_1,\omega_2)$ and uncorrelated pairs of single photoemission events $P_{\mathbf{k}}(\omega_1) \times P_{-\mathbf{k}}(\omega_2)$, respectively, with line cuts for center and difference frequencies shown in (c). Depicted responses are computed for attractive interactions $U/t = -3$ with shape function broadening $\sigma = 4/t$.}
    \label{fig:singleVsPair}
\end{figure*}

By inspection of the denominators in Eq. (\ref{Eq:p-arpes2}), one can expect that for pair ARPES the largest intensity will be obtained for a given ${\bf k_1, k_2}$ when the energies $\omega_1, \omega_2$ are tuned to the respective energy positions of ARPES removal spectra, giving roughly the similar pattern as that obtained by simply multiplying the two independent ARPES spectral functions for photoemitted electrons with opposite spin. 


We focus only on momentum states lying closest to the chemical potential and consider two-particle removal ARPES spectra for opposite spins and momenta ${\bf k_{1,2}}$ drawn from the six degenerate momentum points $(\pm \pi/2,\pm \pi/2), (\pi,0),$ and $(0,\pi)$. Obtaining the eigenstates for the sectors containing
$N_{electrons} = 8, 7, 6$ allows for the construction of Fermi golden rule pair ARPES spectral functions $D_{\bf k_1,k_2}(\omega_1,\omega_2)$ via Eqs. (\ref{Eq:p-arpes1}) and (\ref{Eq:p-arpes2}), or equivalently via Eq. (\ref{eq:pairARPESshapefns}) upon inclusion of the probe shape functions. We focus on spin-resolved pair photoemission spectra; the spin-agnostic response follows via summing equal- and opposite-spin contributions.

The resulting pair photoemission spectra are shown in Fig. \ref{fig:PairARPES}(d)  for $\mathbf{k}_1 = -\mathbf{k}_2 = (\pi/2, \pi/2)$ and opposite spins $\sigma_1 = \uparrow, \sigma_2 = \downarrow$, for both repulsive and attractive interactions. While both cases show a primary peak at equal pair photoemission energies $\omega_1 = \omega_2$, corresponding to the particle-hole-symmetric Hubbard gap, a key new feature is the emergence of a pair of additional peaks for the attractive Hubbard model, with $\omega_1 + \omega_2 = 0$. These directly probe the pair breaking intermediate state and can be understood as a two-step process: First, a photon breaks a Cooper pair to photoemit an electron, while leaving an unpaired electron with pair breaking energy $2\Delta$ in the sample. (2) Subsequently the second photon photoemits this unpaired electron while removing the extra intermediate state energy from the sample. As final state with two electrons removed from a fully-paired superconductor has the same energy as the initial state, the total energy $\omega_1 + \omega_2$ left in the sample by the photoemission process must equal to zero; the intermediate pair breaking state remains encoded in the energy difference $\omega_1 - \omega_2$. These observations can be confirmed by comparing the pair photoemission response to uncorrelated pairs of single photoemission processes, depicted in Fig. \ref{fig:singleVsPair}.

\begin{figure*}[t]
    \centering
    \includegraphics[width=\textwidth]{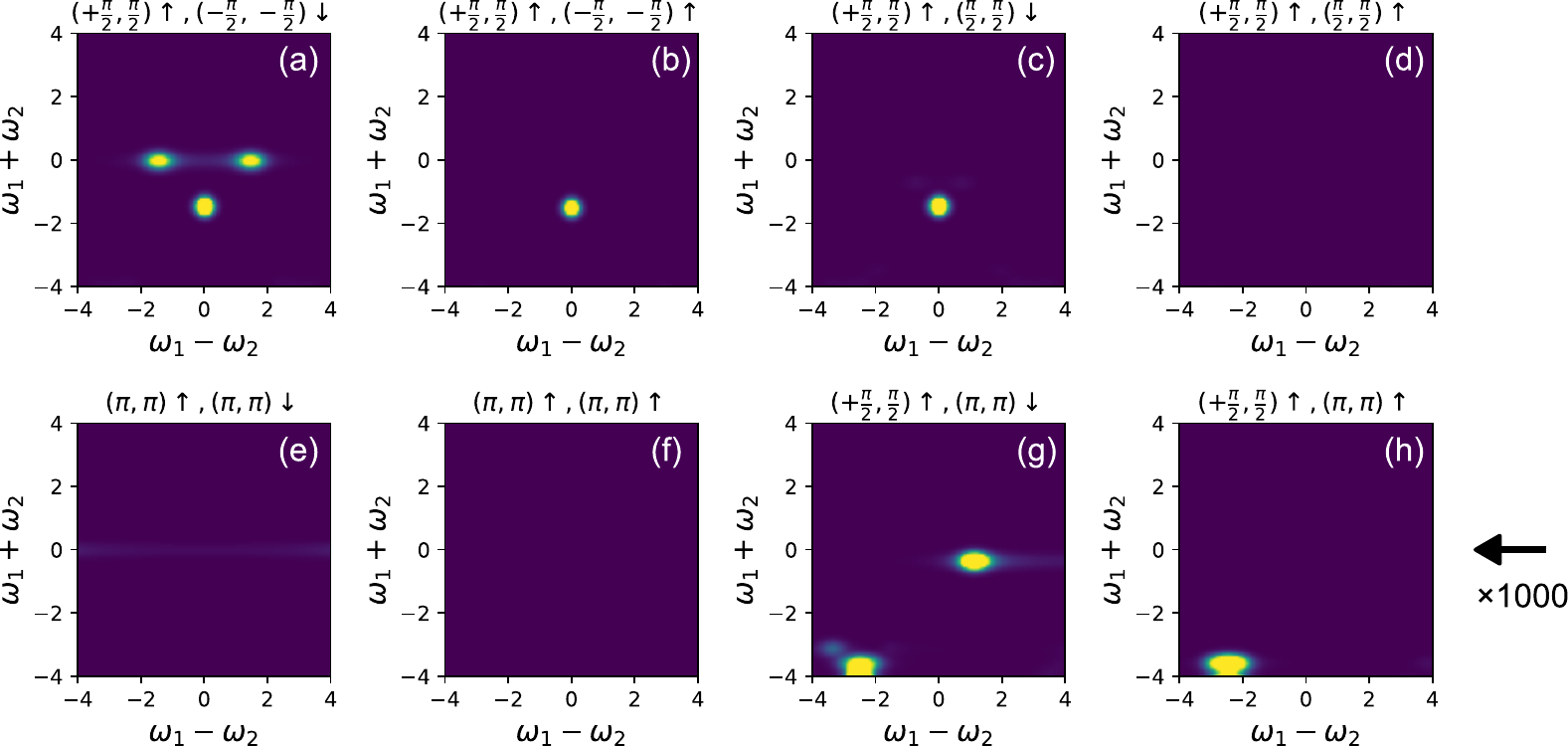}
    \caption{\textbf{Spin and momentum dependence of pair ARPES.} Top (a)-(d) rows depict the pair ARPES response for for the attractive Hubbard model $U/t = -3$, for equal/opposite spin and momentum combinations of the momenta $(\pm \pi/2, \pm \pi/2)$ used in Fig. \ref{fig:PairARPES}. Bottom rows (e)-(h) depict the subdominant pair ARPES response for other momenta (scaled by a factor 1000 with respect to the top row), for equal and opposite emitted spins.}
    \label{fig:differentMomenta}
\end{figure*}



The momentum dependence on each of the fermion momenta as well as the net total momentum ${\bf q} = {\bf k_1}+{\bf k_2}$ and net spin $\sigma=\sigma_1 + \sigma_2$ can reveal further information of the pair wave function.
Fig. (\ref{fig:differentMomenta}) plots pair ARPES for different combinations of photoemitted wavevectors and spins, for $U/t = 3$, as a function of center $\omega_1+\omega_2$ and relative $\omega_1 - \omega_2$ frequencies. As expected for singlet pairing in the attractive Hubbard model, one immediately finds that equal-spin photoemitted electrons [Fig. \ref{fig:differentMomenta}(b)] lack the pair breaking peaks at zero center frequency of the opposite-spin response in Fig. \ref{fig:PairARPES}(d). In contrast, observations of equal-spin pair breaking peaks in the correlated pair ARPES response would be suggestive of triplet pairing instabilities. A similar argument follows for photoemitted pairs of electrons with equal momentum $\mathbf{k}_1 = \mathbf{k}_2$, as shown in Fig. \ref{fig:differentMomenta}(c) and (d); here, an observation of zero-frequency side peaks would be indicative of finite-momentum pairing \cite{PDW}. Pair photoemission for other momenta remains strongly suppressed [Fig. \ref{fig:differentMomenta}(e)-(h)] for attractive interactions.

It is expected that these results will be affected by the finite size and geometry of the small cluster, as well as adding symmetry breaking terms, such as $t'$, that can break momentum degeneracies. For example, the pair-field correlator obtained for the same Hamiltonian on a $4 \times 2$ cluster that breaks $C4$ symmetry, increasing the number of non-degenerate momentum points from 3 in the 8A cluster to 6, has quantitatively the same results for attractive and repulsive $\mid U\mid = 4t$. The largest low frequency contribution is for pair momenta ${\bf q}= (\pi,0)$ and $(\pi/2,\pi)$. By including a negative next nearest hopping $t'=-0.25t$, the low energy pair field correlations are largest for ${\bf q} = (\pi,0)$ and $(0,\pi)$ for $U=4t$, while for $U=-4t$, ${\bf q} =  (0,0)$ still is largest. These effects are larger clusters and different geometries should be further addressed.

Lastly, here we have restricted consideration to zero temperature pair ARPES. One key application of pair ARPES could be to approach ordered phases from high temperature to measure how pair field correlations develop, either through towards a true superconducting transition, or averted by the onset of another competing order, such as charge and/or spin density waves. As these phases all appear to have nearly the same ground state energies in simulations of the Hubbard model, an experimental investigation may provide finer insight into which terms may be missing from the Hubbard model that could formulate a closer contact to materials such as the high temperature superconductors.



In summary, we have presented a theory for pair ARPES whereby two photons produce two photoelectrons detected in coincidence, resolved in both energy and momentum. The corresponding two-particle removal spectra can thus be exploited to determine the effects of electron correlations in a direct way. The calculated pair response for the attractive and repulsive  Hubbard model at half filling for an 8 site Betts cluster shows spectroscopically how prominent ordering tendencies of superconductivity and the net pair momentum and spin can be inferred directly from experiments. 

Experimentally, to make data as close to superconducting pair correlation function as possible, one should try to eject the electron pair from the sample at the same time. In such an “instantaneous event”, two photons eject two electrons in an “interacting volume”, for examples a Cooper pair in a superconductor, or a pair in a Mott insulator that are sufficiently entangled. In a Cooper pair, this means two electrons within the superconducting coherence volume. In a Mott insulator, assuming that Hubbard model is a reasonable starting point, this means two electrons not far from each other so that a cascade of local interactions can entangle the electrons. Our theoretical calculation was carried out in a small cluster such that the entanglement is naturally strong. 

Such pair photoemission is an experiment with many technical challenges. However, several recent technological advances make it realistic. The first is the emergence of much improved and suitable light sources, such as UV lasers, high harmonic generation, free electron lasers, and photon focusing schemes. Photons within a very short pulse, such as tens of femtoseconds, can be considered as identical and instantaneous within time of flight spectrometers having picosecond resolution. The second is the time-of-flight (TOF) based three dimensional ARPES platform, such as the momentum microscope and its spin filtered variant. The third is the development of two-dimensional ultrafast multichannel detectors. With time and through an integration of these important new technologies, enhanced by timing, energy, momentum discrimination schemes and machine learning algorithms to improve the signal to noise ratio, this new spectroscopy may be developed in the near future.

\acknowledgements

Authors would like to thank Frank Marsiglio and Joseph Orenstein for insightful discussions. MZ, JEM, DM, FM, PA, ZXS and TPD acknowledge support for the work from the U.S. Department of Energy (DOE), Office of Basic Energy Sciences, Division of Materials Sciences and Engineering, and through Contract No. DE-AC02-05-CH11231 via the Quantum Materials program (KC2202) (JEM).

\newcommand{\bibtitle}[1]{\textit{#1},}
\newcommand{\bibvol}[1]{\textbf{#1}}



\appendix

\begin{widetext}

\section{Two-Photon Pair Photoemission}

The main text presents a formal expression of the pair photoemission response which assumes negligible backaction between photoemitted electrons and sample electrons. Starting from the second-order perturbative expression
\begin{align}
	D_{\kk_1\kk_2} &= \int\displaylimits_{-\infty}^{t} d\tau_1 d\tau_2 \int\displaylimits_{-\infty}^{\tau_1} d\tau'_1 \int\displaylimits_{-\infty}^{\tau_2} d\tau'_2 ~s(\tau_1) s(\tau'_1) s(\tau_2) s(\tau'_2)  \int d\r_1 d\r'_1 d\r_2 d\r'_2 \int d\x_1 d\x'_1 d\x_2 d\x'_2~ \phi_{\kk_1}^\star(\x_1) \phi_{\kk_2}^\star(\x'_1) \phi_{\kk_2}(\x'_2) \phi_{\kk_1}(\x_2) \notag\\
		&\times e^{i\Wph(\tau_1+\tau'_1 - \tau_2 - \tau'_2)} ~g^\star(\r_1) g^\star(\r'_1)  g(\r'_2) g(\r_2)  \sum_{\sigma_1\sigma'_1 \nu} \sum_{\sigma_2\sigma'_2 \nu'} \left< \PhiS_{\sigma_1}^\dag(\r_1,\tau_1) \PsiF_{\sigma_1}(\r_1,\tau_1) \PhiS_{\sigma'_1}^\dag(\r'_1,\tau'_1) \PsiF_{\sigma'_1}(\r'_1,\tau'_1) \right. \notag\\
        &\times \left. \PsiF_{\nu}^\dag(\x_1,t) \PsiF_{\nu'}^\dag(\x'_1,t) \PsiF_{\nu'}(\x'_2,t)  \PsiF_{\nu}(\x_2,t)  \PsiF_{\sigma'_2}^\dag(\r'_2, \tau'_2) \PhiS_{\sigma'_2}(\r'_2,\tau'_2) \PsiF_{\sigma_2}^\dag(\r_2, \tau_2) \PhiS_{\sigma_2}(\r_2,\tau_2) \right>
\end{align}
Neglecting backaction on the sample permits a decomposition
\begin{align}
	D_{\kk_1\kk_2} &= \sum_{\sigma_1\sigma'_1 \nu} \sum_{\sigma_2\sigma'_2 \nu'} \int\displaylimits_{-\infty}^{t} d\tau_1 d\tau_2 \int\displaylimits_{-\infty}^{\tau_1} d\tau'_1 \int\displaylimits_{-\infty}^{\tau_2} d\tau'_2 ~s(\tau_1) s(\tau'_1) s(\tau_2) s(\tau'_2) e^{i\Wph(\tau_1+\tau'_1 - \tau_2 - \tau'_2)} \notag\\
 &\times \int d\r_1 d\r'_1 d\r_2 d\r'_2 ~g^\star(\r_1) g^\star(\r'_1)  g(\r'_2) g(\r_2) \left< \PhiS_{\sigma_1}^\dag(\r_1,\tau_1) \PhiS_{\sigma'_1}^\dag(\r'_1,\tau'_1) \PhiS_{\sigma'_2}(\r'_2,\tau'_2) \PhiS_{\sigma_2}(\r_2,\tau_2) \right> \notag\\
 &\times \int d\x_1 d\x'_1~ \phi_{\kk_1}^\star(\x_1) \phi_{\kk_2}^\star(\x'_1)~ \left< \PsiF_{\sigma_1}(\r_1,\tau_1) \PsiF_{\sigma'_1}(\r'_1,\tau'_1) \PsiF_{\nu}^\dag(\x_1,t) \PsiF_{\nu'}^\dag(\x'_1,t) \right> \notag\\
 &\times \int d\x_2 d\x'_2~ \phi_{\kk_2}(\x'_2) \phi_{\kk_1}(\x_2)~ \left< \PsiF_{\nu'}(\x'_2,t)  \PsiF_{\nu}(\x_2,t)  \PsiF_{\sigma'_2}^\dag(\r'_2, \tau'_2) \PsiF_{\sigma_2}^\dag(\r_2, \tau_2) \right>
\end{align}
Suppose now that the sample electrons near the Fermi energy are confined to a single valence band $\PhiS_{\sigma}(\r,\tau) \approx \frac{1}{\sqrt{L}} \sum_{\kP} u_{\kP}(\r) \C{\kP\sigma}(\tau) e^{i\kP\rP}$ with Bloch function $u_{\kP}(\r)$. Expanding the photo-emitted electron fields in a plane wave basis, and Fourier transforming the detector form factors $\phi_{\kk}(\x) = \int d\p e^{-i\p\x} \phi_{\kk}(\p)$, one obtains
\begin{align}
	D_{\kk_1\kk_2} &= \sum_{\sigma_1\sigma'_1 \nu} \sum_{\sigma_2\sigma'_2 \nu'} \int\displaylimits_{-\infty}^{t} d\tau_1 d\tau_2 \int\displaylimits_{-\infty}^{\tau_1} d\tau'_1 \int\displaylimits_{-\infty}^{\tau_2} d\tau'_2 ~s(\tau_1) s(\tau'_1) s(\tau_2) s(\tau'_2) e^{i\Wph(\tau_1+\tau'_1 - \tau_2 - \tau'_2)} \notag\\
 &\times \int d\k d\k' d\q~ \left< \CD{\kP,\sigma_1}(\tau_1) \CD{\qP-\kP,\sigma'_1}(\tau'_1) \C{\qP-\kP',\sigma'_2}(\tau'_2) \C{\kP',\sigma_2}(\tau_2) \right> \notag\\
 &\times \int_{\textrm{BZ}} d\r_1 d\r'_1 d\r_2 d\r'_2  ~g^\star(\r_1) g^\star(\r'_1)  g(\r'_2) g(\r_2)~ u^*_{\kP}(\r_1) u^*_{\qP-\kP}(\r'_1) u_{\qP-\kP'}(\r'_2) u_{\kP'}(\r_2) \notag\\
  &\times \int d\p~ \phi_{\kk_1}^\star(\q-\p) \phi_{\kk_2}^\star(\p)~ \left< \PsiF_{\sigma_1}(\k,\tau_1) \PsiF_{\sigma'_1}(\q-\k,\tau'_1) \PsiF_{\nu}^\dag(\q-\p,t) \PsiF_{\nu'}^\dag(\p,t) \right> \notag\\
  &\times \int d\p'~  \phi_{\kk_2}(\p') \phi_{\kk_1}(\q-\p') ~  \left< \PsiF_{\nu'}(\p',t)  \PsiF_{\nu}(\q-\p',t)  \PsiF_{\sigma'_2}^\dag(\q-\k', \tau'_2) \PsiF_{\sigma_2}^\dag(\k', \tau_2) \right>
\end{align}
Rewriting the third line in terms of matrix elements recovers the expression presented in the main text.

\section{Spectral representation of Pair ARPES}

A useful representation of pair ARPES that accounts for the probe shape function follows from a spectral decomposition of Eq. (\ref{eq:CARPES})
\begin{align}
    &D^{(0)}_{k_1,k_2}(\omega_1,\omega_2) = \sum_n \left| \int\displaylimits_{-\infty}^{\infty} dt \int\displaylimits_0^\infty dt' \right. ~ s(t) s(t-t')  \notag\\
    &\times ~\sum_m \left[ \vphantom{\frac{1}{2}} \right. \bra{n} \C{\k_1} \ket{m} \bra{m}{\C{\k_2}}\ket{0} e^{i \omega_1 t'} - \left. \bra{n} \C{\k_2} \ket{m} \bra{m} \C{\k_1} \ket{0} e^{i \omega_2 t'} \right] \left. e^{i[(\E_n - \E_0 - \omega_1 - \omega_2) t - (\E_n - \E_m) t']} \right|^2
\end{align}
with $s(t)$ the Gaussian shape functions. The time integrals can now be evaluated and one arrives at
\begin{equation}
    D^{(0)}_{k_1,k_2}(\omega_1,\omega_2) = \sum_n \left| \sum_m \left[ \bra{n} \C{\k_1} \ket{m} \bra{m}{\C{\k_2}}\ket{0} X_{nm}(\omega_1,\omega_2)  \right.\right. 
    -~ \bra{n} \left.\left. \C{\k_2} \ket{m} \bra{m} \C{\k_1} \ket{0} X_{nm}(\omega_2,\omega_1) \right] \vphantom{\frac{1}{2}} \right|^2   \label{eq:pairARPESshapefns}
\end{equation}
where
\begin{equation}
    X_{nm}(\omega_1,\omega_2) = \sigma^2 e^{-\frac{\sigma^2}{4} \left( \omega_1 + \omega_2 + \E_0 - \E_n \right)^2}
    \left\{ \frac{1}{2} e^{-\frac{\sigma^2}{4} \left( \omega_2 - \omega_1 + \E_0 + \E_n - 2\E_m \right)^2}  
     + \frac{i}{\sqrt{\pi}} \mathcal{D}\left[\frac{\sigma}{2} \left( \omega_2 - \omega_1 + \E_0 + \E_n - 2\E_m \right)\right] \right\}
\end{equation}
Here, $\sigma$ is the Gaussian broadening of the shape function and $\mathcal{D}(x)$ denotes the Dawson function, which is related to the error function $\mathcal{D}(x) = \sqrt{\pi/4} e^{-x^2} \textrm{erfi}(x)$.
\end{widetext}

\end{document}